# PRISMATIC SOFT CUBES

K. KOCSIS

**ABSTRACT**

Soft cells are shapes without sharp corners that can fill the space without gaps and overlaps [2]. A sharp corner is a point on the surface of the solid through which no smooth curve passes. In the paper introducing the concept of soft cells [2], the authors proved that there exists an algorithm that can soften tilings consisting of convex polyhedra, preserving the lattice points and combinatorial structure of the original tiling. Although the algorithm guarantees (with a few restrictions) that there exists a soft tiling that is combinatorially equivalent to the convex polyhedral tiling, the proof does not address how to find all such tilings. For a polyhedral tiling based on a truncated octahedral cell, paper [3] shows how to find all soft tilings for a fixed symmetry group. In this paper, we extend this method and apply it to the cubic lattice, imposing only natural conditions, rather than symmetry constraints. The natural conditions being, the directions of edge half-tangents of the tiling are restricted to lattice directions, and the edges of the tiling are planar. This results in 26 soft cubic cells with different geometries. A total of 68 fundamental domains can be created from the cells, which can be classified into 8 groups based on their lattice symmetry. The paper also presents an algorithmic process (with a corresponding program in language Python) for classifying the 26 non-equivalent geometric cell types.

## 1. INTRODUCTION

A tiling is a covering of space by bounded solids (tiles) without gaps and overlaps. A node of the tiling is a point $P$ in space, at which $\mathrm{n(P)}$ tiles meet, and for all points $P'$ in a small spherical neighbourhood of $P$ one has $\mathrm{n(P')} < \mathrm{n(P)}$, such that $n(P)$ is locally maximal. Considering the cubic lattice tiling $n = 8$ tiles meet at each node. Several other polyhedral tilings are known, i.e. the truncated octahedron, where $n = 4$ tiles meet at each node. A bounded solid is soft, if it has no sharp corners, that is at any surface point there is at least one smooth surface curve [2]. A tiling is soft if each tile in the tiling is soft. If the tiles are identical, they are called soft cells [2].

For a broad class of polyhedral tilings, there exists a soft tiling which agrees to first order with the polyhedral tiling, that is the nodes and the face structure are identical [2]. A specific example is known for the cubic lattice [1].

Considering the truncated octahedron tiling, it was proven in paper [3], that there exist exactly 2 soft tilings that share the full symmetry group of the original polyhedral tiling and agree to first order with it.

We now consider softenings of the standard cubic lattice. The tile is the unit cube, and the nodes of the tiling are the lattice points with integer coordinates in three-dimensional space. This paper shows that if the half-tangent vectors of the edges are restricted to lattice directions, and the edges of the tiles are planar, there exist exactly 26 soft tilings that agree to first order with the cubic lattice. No further constraints (i.e. symmetry) are imposed. It is shown that none of the 26 soft tilings share the full symmetry group of the cubic lattice.

We also aim the geometric realization of these soft tilings and the soft cells and fundamental domains associated with them. A symmetry classification is also provided.

To prove the statement above, we establish three auxiliary theorems leading to Theorem 4, which contains the main result. We prove that the 26 soft tilings are necessarily prismatic. Thus, there exists a direction of projection for which all perpendicular cross-sections form a square lattice. If the edge half-tangents are not restricted to lattice directions and/or the edges need not to be planar, there may exist more prismatic soft cubes. The 26 soft tilings are described with the location of nodes and the unit vectors of edge half-tangents (second order description). Higher-order deviations (shape of edges, shape of faces) are shown for illustrative purposes only.

Theorem 4, which states the main result (existence of the 26 soft tilings), is proved using an algorithm. In addition to describing the algorithm, we also present its implementation. We prove that only the cell published in paper [1] satisfies the following two conditions: the tiling is face-to-face and the symmetry group of the tiling only contains translation in addition to the symmetry group of the tile (Figure 1). If the latter condition is disregarded there is exactly 25 additional soft tilings.

We distinguish between monohedral tilings that fill the space with only translations and/or rotations (achiral tiling), and monohedral tilings that need their mirror images to fill the space (chiral tiling). From the 26 soft cells, 68 tilings can be constructed that differ from the cell published in [1] only at third order. Based on the metric of the tiling lattice, soft tilings may correspond to space groups belonging to the triclinic, monoclinic, orthorhombic, and tetragonal crystal systems (Figures 26-27) [4]. Theorem 5 summarizes these statements.

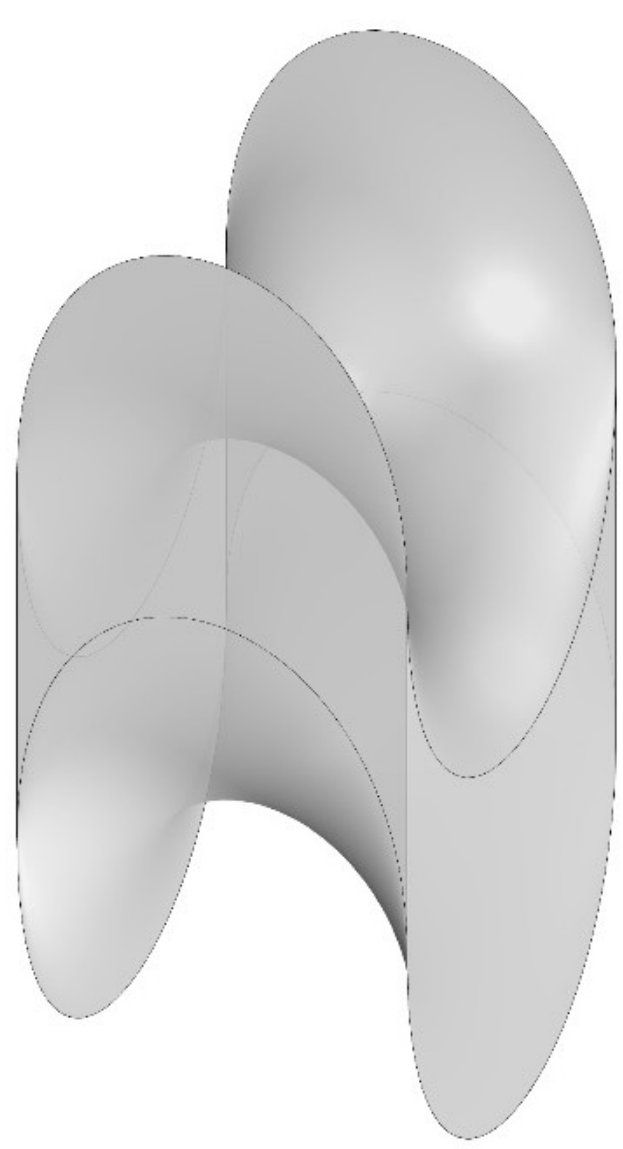

FIGURE 1. Soft cell of the soft tiling that agrees to first order with the cubic lattice published in [1].

### 1.1. KEY DEFINITIONS

The following definitions are taken verbatim from Definitions 1–3 in [3].

To better understand soft tilings and their relationship to natural structures, we introduce below a five-level classification of polyhedral patterns.

**Definition 1.** Let $M$ be a polihedric tiling. Then, the zeroth-order description of $M$ is the combinatorial structure of $M$. The $i$-th order description ($i$=1,2,3,4) of $M$ contains the ($i$-1)th order description and the following additional features:

- $i$=1: the location of nodes,
- $i$=2: the unit vectors of edge half-tangents,
- $i$=3: the shape of edges,
- $i$=4: the shape of faces.

We associate the $i$-th order description with the symmetry group $\Gamma_\mathrm{i}$ and we have $\Gamma_\mathrm{i} \leq \Gamma_\mathrm{j}$ if $i \geq$ j. Definition 1 immediately admits a convenient new definition of softness:

**Definition 2.** Let $M$ be a polyhedric tiling with smooth edges and faces. A cell of $M$ is soft if each node has at least two half-tangents unit vectors $\mathbf{u_1}, \mathbf{u_2}$ such that $\mathbf{u_1} \cdot \mathbf{u_2} = \mathbf{-1}$. A tiling is soft if each cell in the tiling is soft.

**Definition 3.** We assume that the polyhedric tiling $M$ is defined to first order, so the symmetry group $\Gamma_1$ is known and we also know the symmetry group $\Gamma_2 \leq \Gamma_1$ associated with the second order description. We assume that at each node N cells and K edges meet, so we have K unit vectors as half-tangents, which we call the *nodal set* of M. From the nodal set we can construct N different subsets containing half-tangents, belonging to the N cells. We will refer to these sets of vectors as the N *vertex sets* of M and we denote the size of the vertex sets by $v_i, i = 1,2, \dots N$.

Using the notions in Definition 3, the steps of the Extended Edge Bending (EEB) algorithm (and the modified E*EB algorithm) are the following:

(1) We identify a maximal set of unit half tangents $\boldsymbol{u_i}, i = 1,2, \dots f$ which are not related by any transformation of the group $\Gamma_2$.
(2) Using the set $\boldsymbol{u_i}, i = 1,2, \dots f$ as inputs, we apply symmetry transformations in $\Gamma_2$.to construct the nodal set $\boldsymbol{u_i}, i = 1,2, \dots n, n \geq f$ of M.
(3) We pick a pair of half tangents $\mathbf{u_i}, \mathbf{u_j}$ from the nodal set and start writing a list:

$$\mathbf{u_i} \cdot \mathbf{u_j} = \mathbf{-1}$$

We continue picking pairs and add the corresponding equation to list if the equation does not agree with any previous equation and does not contradict any previous equation. We continue this process until we have at least one equation in every vertex set. At this point, we denote the number of equations by $E$ and we call this system of equations a *complete set of softening equations* of M. The same tiling may have several complete sets of softening equations.
(4) We let the vectors $\boldsymbol{u_i}, i = 1,2, \dots f$ of the fundamental domain independently run over the boundary of unit sphere. Since all vectors in the nodal set can be computed via transformation matrices of the fundamental domain, this operation will transform the softening equations into $E$ equations on the sphere in 2 variables.
(4*) We let the vectors $\boldsymbol{u_i}, i = 1,2, \dots f$ of the fundamental domain independently run over the 6 directions given by the standard cubic lattice. All edges in M′ must be $C^1$ planar curves, that is, no two adjacent vertices admit deviation vectors.
(5) If there are additional, prescribed constraints on the geometry of the cells (e.g. planar faces) then we also solve the corresponding equation systems.
(6) We combine the solutions guaranteeing soft geometry with those guaranteeing additional constraints.

The existence of a solution of any of these systems is a necessary (but not sufficient) condition for the existence of a soft tiling M′ which is equivalent to first order to M, has symmetry group $\Gamma_2$ and obeys the prescribed additional geometric constraints. Symmetry is not guaranteed since we may have been able to compute the nodal set by applying only a subset of $\Gamma_2$.

**Definition 4.** Let $M$ be a convex tiling and let $V_i$ be the $i$-th node of $M$ with nodal degree $n_i$. Let the set of half tangents of the polyhedral edges (nodal set) at $V_i$ be denoted by $H_i = \{h_{i,j}\}, j = 1,2, \dots n_i$. Let M′ be a soft tiling produced by the Edge Bending Algorithm, with $M$ as the initial input and let the set of half tangents of $M'$ at $V_i$ be denoted by $H_i'$. If $\forall j: h_{i,j} = h_{i,j}'$ then $H_i' = H_i$ and, using Definition 1 we say that $M$ and $M'$ agree to second order. If $H_i' \subseteq H_i$ then we say that $M'$ and $M$ *partially agree* to second order. If, in addition, all edges in $M'$ are planar then we say that $M'$ is a *parallel tiling*.

**Remark 1.** *In this paper, instead of Step 4 we use Step 4* in the EEB Algorithm and* $\Gamma_2 = \{I\}$.

## 2. SOFTENING THE NODES OF THE CUBIC LATTICE

Consider a cubic lattice where K=6 edge half-tangent unit vectors meet at each node. These half-tangent vectors form the nodal set $H_i = a, b, c, d, e, f$. (Figure 2)

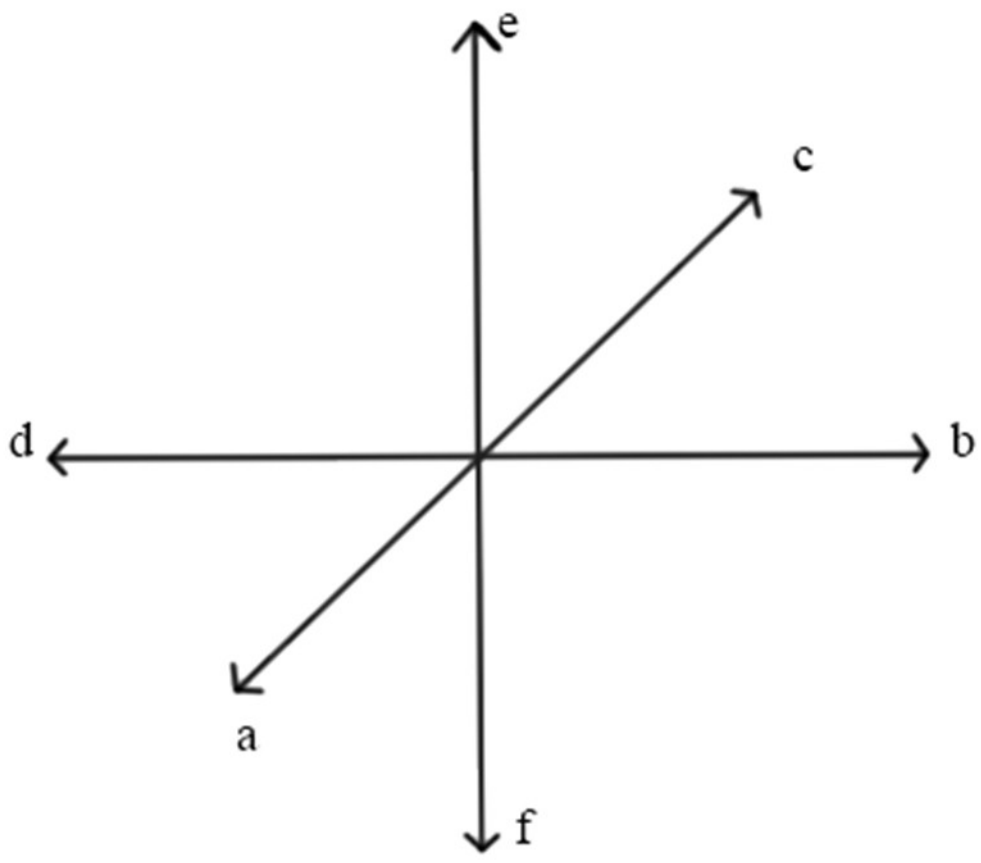

FIGURE 2. K=6 edge half-tangent unit vectors meet at each node of the cubic lattice. These half-tangent vectors form the nodal set $H_i = a, b, c, d, e, f$.

We apply the softening algorithms introduced in [2] and [3] to the case where the translation lattice is a cubic lattice, the lattice points serve as the nodes of a tiling, and the tiling cells are combinatorially equivalent to a cube. At each node, eight corners must meet, locally, each of these corners are bounded by the three edges incident to the node and the three faces spanned by the corresponding edges. It is necessary to guarantee the softness of the eight corners meeting at each node. Since the corners occurring at a node are congruent to the corners at the vertices of the tile, the softness of all corners required for the softness of the cell must be ensured individually.

For each corner, among its three edges, two must always have mutually opposite tangents, so that two of the edges of the corner form a smooth curve passing through the node. The following procedure searches for configurations satisfying this system of conditions.

**Theorem 1.** *The complete sets of softening equations of parallel soft tilings partially agree to second order with the cubic lattice admit 12 solutions.*

*Proof*

Faces incident at a node are defined by pairs of half-tangent vectors, and corners by triples of half-tangent vectors. Using the notation of Figure 2, the faces are represented by the pairs $ea, eb, ec, ed, fa, fb, fc, fd, ab, bc, cd, da$, whereas the corners are represented by the triples $eab, ebc, ecd, eda, fab, fbc, fcd, fda$. The **tiling cell is soft** if, in each of the eight corners listed above, there exists a pair of edges determining opposite tangents. Softness is therefore ensured by a system of equations, whose selection is carried out as described below.

Let $A_i$ and $B_i$ be two subsets of the nodal set $H_i$. Elements of $A_i$ and $B_i$ will form opposite half-tangent vectors ($H_i = A_i \cup B_i$). The general case is obtained by specifying three pairs of edges whose half-tangents, after softening, become complementary half-lines of distinct spatial lines. We do not exclude the possibility that two or all three of these lines coincide. In the general softening algorithm presented in [2], coincidence of these lines is assumed; therefore, the present procedure may be regarded as a generalization of that method. The choice of $A_i$ and $B_i$ is arbitrary, however we assume that each subset contains a tangent pair corresponding to complementary half-tangents, denoted by $ac$ and $bd$. These edge half-tangent pairs do not determine a face in any corner and therefore yield equations that are irrelevant in softening.

Let $A_i = \{a, c, e\}$ and $B_i = \{b, d, f\}$. Then the set pairs between elements of $A_i$ and $B_i$ is the Cartesian product $A_i \times B_i$:

$$A_i \times B_i = \{(x, y) \mid x \in A_i, y \in B_i\} \tag{1}$$

The pairs explicitly:

$$\mathrm{A_i \times B_i = \{(a, b), (a, d), (a, f), (c, b), (c, d), (c, f), (e, b), (e, d), (e, f)\}}. \tag{2}$$

Let $\mathrm{C_{i,j} \subset A_i \times B_i},\ \mathrm{j} = 1,2, \dots ,6$, where $C_{i,j}$ is a three-element subset of $A_i \times B_i$ corresponding to a bijection between $A_i$ and $B_i$, i.e. each element of $A_i$ and $B_i$ appears exactly once, and $j$ is the index of the bijection. Since the number of bijections is $3! = 6$, the bijections are:

$$C_{i,1} = \{(a, b), (c, d), (e, f)\}, \tag{3}$$
$$C_{i,2} = \{(a, b), (c, f), (e, d)\}, \tag{4}$$
$$C_{i,3} = \{(a, d), (c, b), (e, f)\}, \tag{5}$$
$$C_{i,4} = \{(a, d), (c, f), (e, b)\}, \tag{6}$$
$$C_{i,5} = \{(a, f), (c, b), (e, d)\}, \tag{7}$$
$$C_{i,6} = \{(a, f), (c, d), (e, b)\}. \tag{8}$$

Elements of $C_{i,j}$ are equations of the form $a \cdot b = -1$, $C_{i,j}$ sets are incomplete sets of softening equations. To soften the cubic cell, at least four equations of the form $a \cdot b = -1$ are required. Since, a pair of half-tangent vectors can determine a face in at most two corners, therefore, even in the most favourable case, three half-tangent pairs can ensure softness in no more than six corners. To complete the sets of softening equations, $C_{i,j}$ must be extended by an additional pair of half-tangents defining an equation from the same subset, either $A_i$ or $B_i$. The number of half-tangent pairs in a subset is $\binom{3}{2} = 3$. The pairs in $A_i$ are $\{a, c\}$, $\{a, e\}$, $\{c, e\}$. Since we assumed that the pair $ac$ corresponds to complementary half-tangents, therefore declared its irrelevance, the fourth equation can be combined with each of the six existing equation systems $C_{i,j}$, $j = 1,2,\dots,6$ in $6 \cdot 2 = 12$ distinct ways. The generation of the equation systems is summarized in Table 1, and their solutions are illustrated graphically in Figure 3.

■

| | $A$ $(a,e)$ | $B$ $(c,e)$ |
|---|---|---|
| 1 $C_{i,1} = \{(a,b),(c,d),(e,f)\}$ | $(a,b),(c,d),(e,f),(a,e)$ | $(a,b),(c,d),(e,f),(c,e)$ |
| 2 $C_{i,2} = \{(a,b),(c,f),(e,d)\}$ | $(a,b),(c,f),(e,d),(a,e)$ | $(a,b),(c,f),(e,d),(c,e)$ |
| 3 $C_{i,3} = \{(a,d),(c,b),(e,f)\}$ | $(a,d),(c,b),(e,f),(a,e)$ | $(a,d),(c,b),(e,f),(c,e)$ |
| 4 $C_{i,4} = \{(a,d),(c,f),(e,b)\}$ | $(a,d),(c,f),(e,b),(a,e)$ | $(a,d),(c,f),(e,b),(c,e)$ |
| 5 $C_{i,5} = \{(a,f),(c,b),(e,d)\}$ | $(a,f),(c,b),(e,d),(a,e)$ | $(a,f),(c,b),(e,d),(c,e)$ |
| 6 $C_{i,6} = \{(a,f),(c,d),(e,b)\}$ | $(a,f),(c,d),(e,b),(a,e)$ | $(a,f),(c,d),(e,b),(c,e)$ |

**Table 1.** Complete sets of softening equations for the cubic lattice. Each element of the form $(a,b)$ represents the equation $a \cdot b = -1$ , where $a$ and $b$ are half-tangent unit vectors at the nodes of the tiling. $\mathrm{C_{i,j}} \subset \mathrm{A_i} \times \mathrm{B_i}, \mathrm{j} = 1,2,\dots,6$ : incomplete sets of softening equations as defined in equations (1)-(8). $(a,e)$ and $(c,e)$: 4th equation completing $\mathrm{C_{i,j}}$ sets.

**Theorem 2.** *Among the solutions of the twelve complete sets of softening equations, exactly four equivalent solutions guarantee the softness of all eight corners at a node.*

*Proof*

It remains to determine which equation systems yield identical solutions or rotated copies of the same solution. It can be verified that there is only one solution that ensures the softness of all eight incident cells. The node configuration of the soft cubic cell described in [1] is as follows:

$$2\mathrm{B} = 5\mathrm{A} = 6\mathrm{A} = 4\mathrm{B}$$

The remaining solutions are admissible at the level of the lattice node but are invalid with respect to the corners. Among these, we identify the following two types:

$$1\mathrm{A} = 2\mathrm{A}$$

$$1B = 3B = 5B = 6B = 3A = 4A$$

The equivalence of the equation systems can be described by the following transformations:

$$1\mathrm{A} \equiv 2\mathrm{A}$$

$$1\mathrm{B} \equiv 6\mathrm{B}$$

$$3\mathrm{A} = 4\mathrm{A} = 1\mathrm{B} \cdot \mathrm{T_y}(\pi)$$

$$3\mathrm{B} = 5\mathrm{B} = 1\mathrm{B} \cdot \mathrm{T_y}\left(\frac{\pi}{2}\right)$$

$$2\mathrm{B} \equiv 4\mathrm{B}$$

$$5A = 6A = 2B \cdot T_y\left(\frac{\pi}{2}\right)$$

$T_y\left(\frac{\pi}{2}\right)$ and $T_y(\pi)$ matrices:

$$T_y\left(\frac{\pi}{2}\right) = \begin{pmatrix} 0 & 0 & 1 \\ 0 & 1 & 0 \\ -1 & 0 & 0 \end{pmatrix}$$

$$T_y(\pi) = \begin{pmatrix} -1 & 0 & 0 \\ 0 & 1 & 0 \\ 0 & 0 & -1 \end{pmatrix}$$

Configuration $1A$ (Figure 4) does not provide softness for the corners $daf$ and $ebc$, while configuration $1B$ (Figure 5) does not provide softness for the corners $fbc$ and $ead$.

■

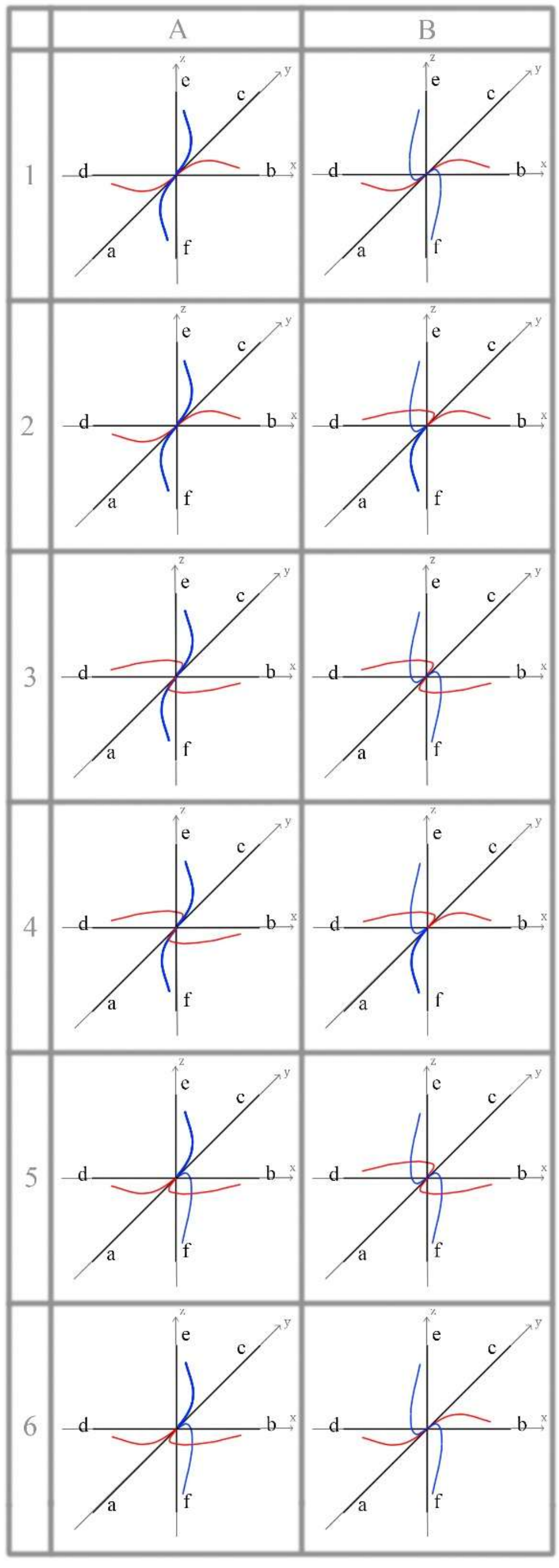

FIGURE 3. Solutions of the complete sets of softening equations for the cubic lattice, using the notation of Table 1. Red edges lie in the $[x, y]$-plane, blue edges lie in the $[y, z]$-plane.

It follows that a parallel soft, gap-free and overlap-free tiling that partially agrees to second order with the cubic lattice can be constructed only if the nodal softening conditions listed in Table 1 are satisfied. These conditions yield a soft cell if and only if, at each node, every incident cell possesses a pair of edges whose tangent directions are opposite. The only soft cube that can be generated only by translations of the nodes is the one given in [1].

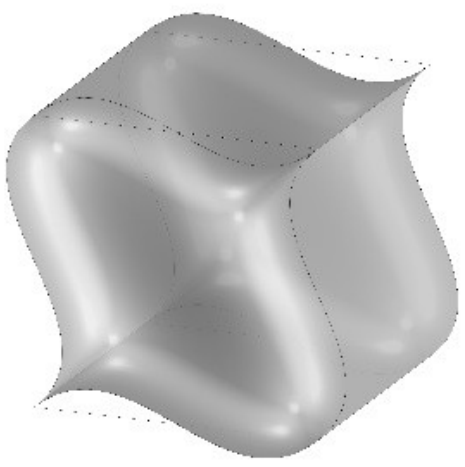

FIGURE 4. Non-soft cell $1A = 2A$, using the notation of Table 1.

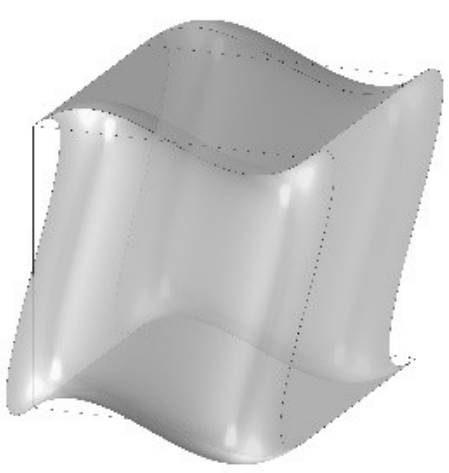

FIGURE 5. Non-soft cell $1B = 3B = 5B = 6B = 3A = 4A$, using the notation of Table 1.

**Theorem 3.** *Parallel soft tilings that fill the space without gaps and overlaps and partially agree to second order with the cubic lattice have prismatic structure. Thus, there exists a direction of projection for which all perpendicular cross-sections form a square lattice.*

*Proof*

In Theorem 2 we establish that the half-tangent unit vectors $a, b, c, d, e, f$ at a node ensure the softness of the associated eight corners if and only if they form three coincident pairs, with the distinct pairs being mutually opposite. It follows that, for a soft monohedral tiling with cubic combinatorics, there exists at each node a line whose two opposite half-lines determine the tangent directions of all six edges incident to the node. Equivalently, the face angles of the corners attain only the values π and $0$. Since the nodes are equivalent under the symmetry group, the half-tangent vectors at every node of the tiling lie on a line parallel to the line described above.

We distinguish two types of edges, according to whether the two endpoints of a given edge lie on the same such line or on two distinct such lines. In the latter case, the tangents are transversal to the line connecting the endpoints (typically orthogonal), whereas in the former case the two lines coincide. Since the three pairs of nodes determined by the edges of a corner incident to a given node define three non-coplanar segments, at most two edges can belong to the second type. Consequently, there exists an edge whose endpoint tangents are aligned with the line joining its endpoints. By convention, such pairs of nodes are connected by a straight segment. Since two lattice translations and their composition map this edge to three further edges of the same cell, the cell possesses four parallel and congruent edges. The direction of the third lattice translation is parallel to these edges, and its magnitude equals their common length. Therefore, the tiling has a prismatic structure with one free parameter, the length of the straight edge of the cell.

■

### 3. SOFTENING THE CORNERS OF A CUBIC LATTICE

Hereafter, we no longer require every node to be in identical position. From the equation systems listed in Table 1 it follows that the tangent configuration at a node can occur in exactly one way if all eight corners at that node are to be soft. Retaining the nodes of the cubic lattice, all soft tilings can be generated - without imposing any prior symmetry assumptions - by considering combinations of the transforms of the admissible node configuration (Figure 3, configuration $2B$) within a single cell. Namely, the node configuration is varied at each of the eight vertices of a cell, and all possible arrangements are enumerated.

**Lemma 1.** *By fixing one endpoint of an edge and rotating it about that endpoint, the node configuration described in Theorem 2 yields four distinct types of curved edges, classified according to the directions of the half-tangents at the endpoints. These are illustrated in Figure 6. Therefore, six different types of curved faces arise from compositions of the four edge types, while preserving the softness conditions. The six curved face types are shown in Figures 7 and 8.*

**Remark 2.** *Considering Theorem 3, four of the six faces of the cell are planar, therefore, the further classification is determined by the relative configuration of the two remaining opposite curved faces.*

*Proof*

By rotating two node configurations of type $2B$ (Figure 3), it follows that four distinct types of curved edges can arise, classified according to the directions of the half-tangents at their endpoints. The half-tangent directions at the endpoints of an edge are either identical (type 1) or opposite (type 2), the corresponding mirror images of these edges are denoted by 1' and 2', respectively. In the classification of edges, only the endpoint half-tangent directions are prescribed, not the precise geometric shape of the edge.

From the four curved edge types, curved faces are constructed in such a way that the softness conditions are preserved (Figure 8). The faces are labelled by Roman numerals, and their defining edges are specified—following the traversal direction shown in Figure 7 - using the notation introduced in Figure 6.

- A face is composed exclusively of edges of the same type:
  - The directions of the half-tangents at the endpoints coincide along every edge. The edge types forming the face are, in cyclic order 1-1'-1-1'.

- The directions of the half-tangents at the endpoints are opposite along every edge. The edge types forming the face are, in cyclic order 2-2-2-2.
- Mirror image of type II. The edge types forming the face are, in cyclic order 2'-2'-2'-2'.

- Different edge types within a single face:
  - IV. Edges of identical type occur next to each other along the face boundary. The edge types forming the face are, in cyclic order 2'-2'-1'-1.
  - V. Mirror image of type IV. The edge types forming the face are, in cyclic order 2-2-1-1'.
  - VI. Edges of the same type lie opposite each other on the boundary of the face. The edge types forming the face are, in cyclic order 2'-1'-2'-1.

  In faces with mixed edge types (1/1'-2/2'), types 1 and 2 must occur in equal proportion, otherwise, the softness conditions fail.

More curved face types cannot be realized while preserving the softness conditions. We therefore examine the admissible combinations of curved face types.

We assign to each face itself, all other faces, and their rotations about the $z$-axis by $\frac{\pi}{2}$ (I'), π (I''), $\frac{3}{2}\pi$ (I'''). In this manner, the total number of combinations considered is $6 \cdot 6 \cdot 4 = 144$. The number of cases to be examined can be further reduced by taking the symmetries of the faces into account.

■

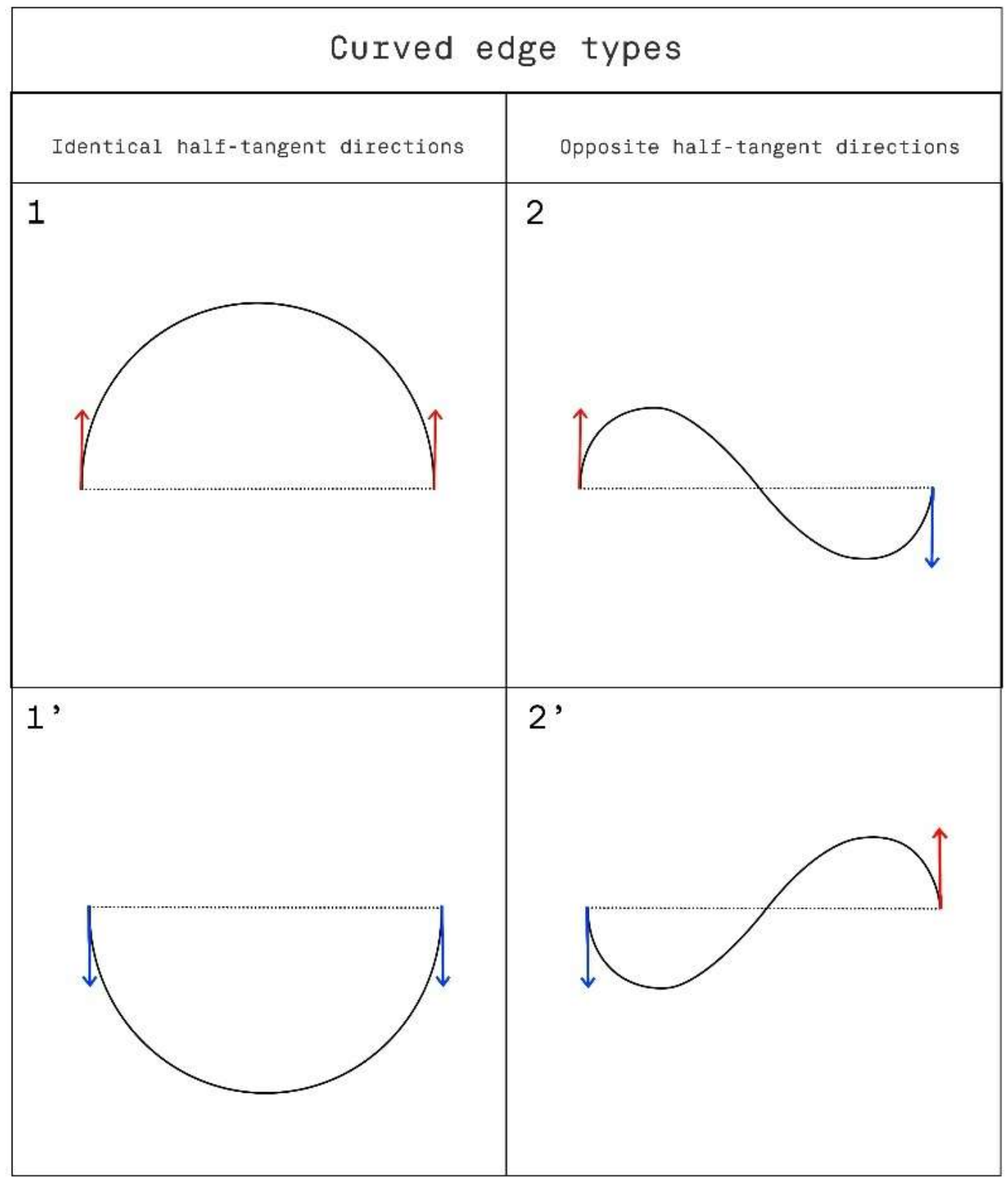

FIGURE 6. Curved edge types classified according to the directions of the half-tangents at their endpoints. The edge segment between the endpoints and their local neighbourhoods is represented in schematic way. Identical tangent directions at the endpoints of the curved edge: 1 and 1’. Opposite tangent directions at the endpoints of the curved edge: 2 and 2’.

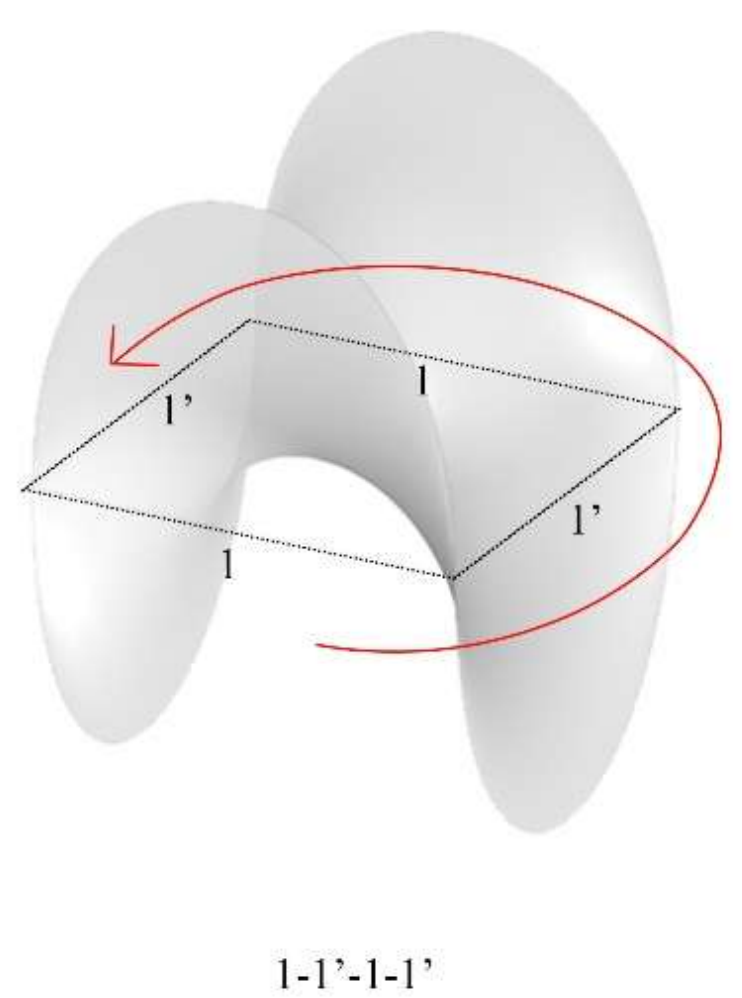

1-1’-1-1’

FIGURE 7. Traversal direction used in the specification of curved face types: anticlockwise. The figure shows curved face type I. Using the notations of Lemma 1 the order of the curved edges: 1-1'-1-1'.

| Curved face types | | | | | |
|---|---|---|---|---|---|
| Edges of identical types | | | Edges of different types | | |
| Identical half-tangent directions | Opposite half-tangent directions | | Identical type edges are adjacent | | Identical type edges are opposite |
| I. | II. | III. | IV. | V. | VI. |
| 1-1’-1-1’ | 2-2-2-2 | 2’-2’-2’-2’ | 2’-2’-1’-1 | 2-2-1-1’ | 2’-1’-2’-1 |

FIGURE 8. Curved face types specified by the curved edge types bounding them, using the notations of Lemma 1. Column (I): identical type edges in one face, identical tangent directions at the endpoints of the curved edges (1-1'-1-1'). Columns (II) and (III): identical type edges in one face, opposite tangent directions at the endpoints of the curved edges (2-2-2-2, 2'-2'-2'-2'). Columns (IV) and (V): different type edges in one face, identical type curved edges are adjacent (2'-2'-1'-1, 2-2-1-1'). Column (VI): different type edges in one face, identical type curved edges are non-adjacent (2'-1'-2'-1).

## 4. SOFT CUBES AND THEIR SYMMETRIES

**Theorem 4.** *There are exactly 26 parallel soft tilings which partially agree to second order with the cubic lattice.*

| | I | II | III | IV | V | VI |
|---|---|---|---|---|---|---|
| I | I-I, I-I' | I-III | | I-IV, I-IV' | | I-VI, I-VI' |
| II | | | | IV-II | | VI-II |
| III | | III-II | III-III | IV-III | | |
| IV | | | | IV-IV, IV-IV' IV-IV'', IV-IV''' | IV-V, IV-V' IV-V'' | IV-VI, IV-VI', IV-VI'', IV-VI''' |
| V | | | | | | |
| VI | | | | | | VI-VI, VI-VI', VI-VI'' |

**Table 2.** 26 soft cells generated from the cubic lattice, specified by curved face types with the notations used in Lemma 1 and Figure 8 (i.e. I-III). I,II,III,IV,V,VI: curved face types.

*Proof*

The method used in the proof of Lemma 1 produces duplicates. These are eliminated by comparison process. Two cells are regarded as identical if one can be mapped onto the other by a rotation, a composition of rotations, or reflection. The method of comparison is not trivial, so a manual method (Algorithm 1) and an algorithm is introduced (Algorithm 2). In the manual analysis, we must determine which of the $\binom{144}{2} = 10296$ pairings need to be checked.

**Algorithm 1.** The steps of the manual algorithm are the following:

1. Let $L_i, i = 1,2,3,4$ be curved face sets containing the sorted curved face types $(I, II, III, IV, V, VI)$:
   - $L_1 = \{I\}$,
   - $L_2 = \{II, III\}$,
   - $L_3 = \{IV, V\}$,
   - $L_4 = \{VI\}$.

   Since mirror images are considered identical, II and III, IV and V are mirror images, hence they are sorted into the same curved face sets $L_2, L_3$ respectively. Two cells need not to be compared if their curved face sets are different (i.e. curved face sets for the I-III cell: $L_1 - L_2$, for the I-IV cell: $L_1 - L_3$).
2. We determine the planar face types generated by the curved face types (Figure 9). Two cells need not to be compared if any of their planar faces are different.
3. Traversal direction and orientation of the planar faces must be checked for cells with identical curved face sets and planar face types.

Thus, the 26 distinct soft cells are obtained manually. For the 26 soft cells, in 9 cases the tiling can be obtained by translation and rotation of a single tile (achiral tiling), whereas in 17 cases a mirror copy (which cannot be mapped onto the original copy) is required (chiral tiling).

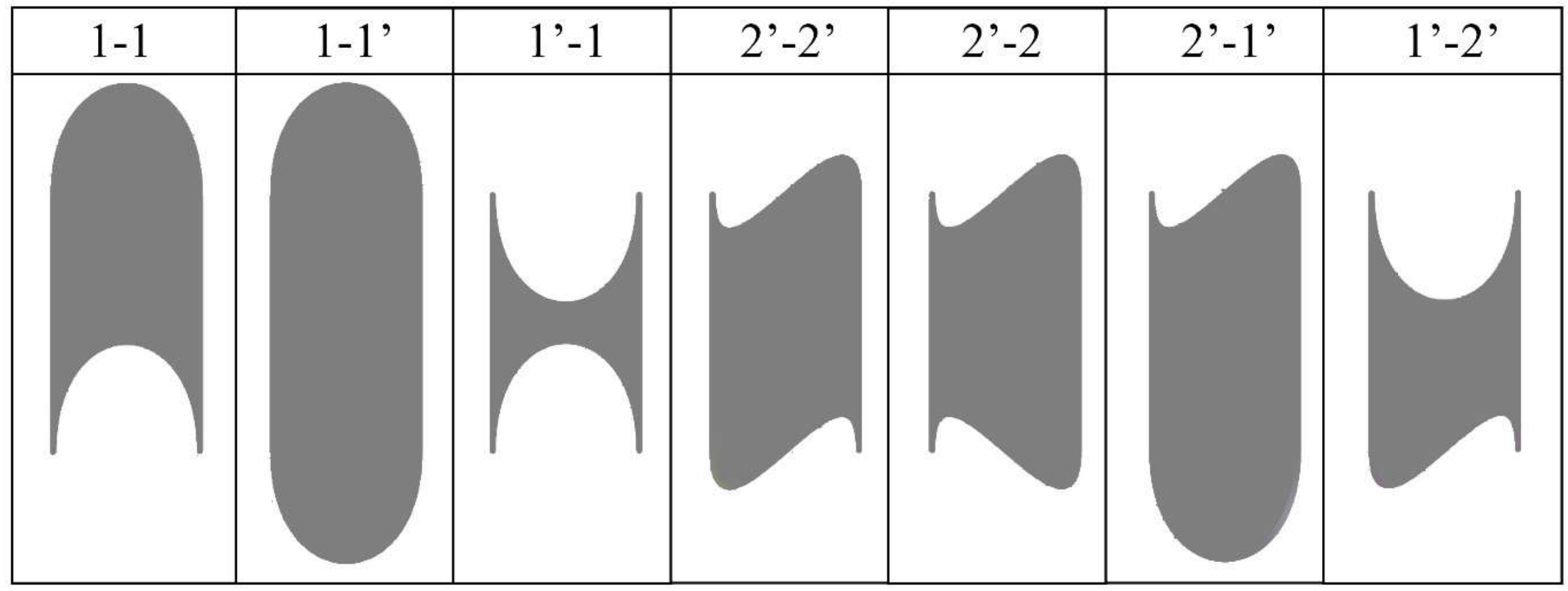

FIGURE 9. Planar face types specified by the curved edge types using the notations of Lemma 1.

**Algorithm 2.** The steps of the algorithm are the following:

1. As an input curved edge types are given with their half-tangent vectors at the endpoints.
   - $\mathrm{A} = [0,0]$ (notation used in Lemma 1: 1')
   - $\mathrm{B} = [1,0]$ (notation used in Lemma 1: 2)
   - $\mathrm{C} = [1,1]$ (notation used in Lemma 1: 1)
   - $\mathrm{D} = [0,1]$ (notation used in Lemma 1: 2')
2. Curved edge sequences are formed while preserving the softness conditions.
3. Curved faces occur from the four-element curved edge sequences. This results 16 curved faces, which are the 6 curved face types in Lemma 1 (I, II, III, IV, V, VI) and their rotations about the $z$-axis by 0, $\frac{\pi}{2}$, π, $\frac{3}{2}$π, where the rotations about the $z$-axis yield different orientations (I', IV',IV'',IV''',V',V'',V''',VI',VI'',VI''').
4. According to Theorem 3 cells arise from pairings of curved faces. Since the algorithm identifies 16 curved faces (instead of 6), the relative rotation of the two curved faces was considered. A list (CELLAK) contains the possible $16 \cdot 16 = 256$ pairings of the curved faces.
5. We define the transformations under which two cells may be mapped onto each other. These are rotations about the $z$-axis by 0, $\frac{\pi}{2}$, π, $\frac{3}{2}$π, and about the $y$-axis by 0, π, and their compositions.
6. We determine the possible combinations of transformations. As an input, the curved face pairs are given, and all transformed instances associated with the curved face pair are obtained as output:
   a) rotations about the $z$-axis by $0, \frac{\pi}{2}, \pi, \frac{3\pi}{2}$ (z)
   b) rotations of a) about the $y$-axis by $0, \pi$ (<u>zy</u>)
   c) rotations of b) about the $z$-axis by $0, \frac{\pi}{2}, \pi, \frac{3\pi}{2}$ (<u>zyz</u>)
   d) rotations of c) about the $y$-axis by $0, \pi$ (<u>zyzy</u>)

7. Given two cells, we apply all combinations of transformations to one, if it can be mapped onto the other by any of these transformations, the function returns true, otherwise false.
8. We collect the distinct cells by initializing an empty list (KULONBOZO=[]). Using the function in Step 7, each cell is compared with all others. The output is a list (KULONBOZO), which contains only cells of distinct geometry.

In the algorithm we do not define mirror transformations, thus, it results the 9 achiral and 17 chiral cells including the mirror copies of the chiral cells. In total $9 + 17 \cdot 2 = 43$ copies.

It is apparent that the number of distinct cases depends on the symmetry groups of the combined faces. The algorithm shows that there are altogether 26 distinct parallel soft cells that partially agree to second order with the cubic lattice and fill space without gaps or overlaps. Table 2 and Figure 10 present the soft cells generated from the face types in a systematic arrangement. Figures 13-16 list the individual soft cubic cells together with their symmetry groups, while Figures 17-21 give the corresponding fundamental domains and their symmetry groups.

■

To analyse the space-filling property of the soft cells, we must determine a domain whose simple translation reproduces the tiling. By applying suitable transformations to a given cell, one can construct a translatable unit, which is a necessary condition for gap-free and overlap-free space filling. This domain is called a period of the tiling. Within a period, translated layers may occur in the z-direction, however, only the top face of the closing element may be attached face-to-face to the bottom face of the initial element. The admissible transformations used in forming a period are reflections, rotations, and their compositions. Owing to the symmetries of the faces, a given cell may admit more than one period.

The minimal size of a period (the smallest number of cells that makes the tiling translatable) is called the fundamental domain of the tiling. In contrast to a period, a fundamental domain may not contain full layers translated in the $z$-direction. In constructing a period, symmetry transformations of the given cell are applied while preserving the face-to-face property of the tiling, thus, each edge must be an edge of all four incident cells. Depending on the symmetries of the curved faces, several distinct periods may be constructed for a given cell.

In the following, we investigate the generation and distinguishability of fundamental domains. We establish the following theorem.

**Theorem 5.** *From 26 soft cubic cells, 68 distinct fundamental domains can be generated, which can be classified into 8 classes.*

*Proof*

As a first step, we outline the method for generating fundamental domains:

1) Construction of translational compatibility in the $x$-direction.
2) Construction of translational compatibility in the $y$-direction.
3) Construction of translational compatibility in the $z$-direction.

Steps 1 and 2 make it straightforward to determine, in each case, how many transformations are required. If two opposite planar faces are not translates of one another, then the cell must be reflected in the plane of that face to obtain a unit that is translatable in the given direction.

Step 3 requires a more detailed analysis. For curved faces, one must determine in how many distinct ways a given cell can be fitted onto a surface formed by prescribed curved faces. It can be verified that a subsequent layer can be attached in two different ways for face type I, in four different ways for face types II and III, in one way for face types IV and V, and in two ways for face type VI. The symmetry of the opposite curved face paired with the given face type determines which of these placements yield genuinely distinct tiling configurations.

Taking this into account, we first construct a period and then examine how far it can be reduced. During the construction, the required face orientation at the point of closure becomes apparent. It follows that the size of the fundamental domain is determined by the individual and combined symmetries of the two curved faces, rather than by the symmetry group of the entire cell.

If the types and orientations of the two opposite curved faces do not coincide, then multiple layers are necessarily required in the $z$-direction. The number of additional layers needed depends on how many transformations are required to stack cells so that the original bottom face reappears, with the same orientation, as the top face of the closing element.

In the present work, we have taken a single cell as the basic unit of construction. The search procedure can also be approached from the side of the tilings themselves by considering, instead of individual cells, the fundamental domains built from them as the primary units. We do not distinguish fundamental domains by their height in the $z$-direction, nor by the edge patterns induced on the planar faces. Under this convention, the fundamental domains can be classified according to their lattice symmetries. For those cell types whose two opposite curved faces belong to different face types or admit multiple mirror realizations in the $x$- and $y$-directions that still produce a valid translation unit, several distinct fundamental domains of equal size can be constructed.

For each cell, one may choose whether the lower or the upper curved face serves as the curved-face component of the fundamental domain. If the lower and upper faces of the cell are congruent, exactly one extension option arises in the $z$-direction; otherwise, two distinct extension options occur. It is therefore necessary to determine in how many ways the cell can be extended in the $x$- and $y$-directions. Face type I requires no extension in the $x$- or $y$-direction. Face types II, III, and VI admit a unique extension in these directions, whereas face types IV and V admit three distinct extensions. Face type IV can be reflected in the planes of its edge pairs 2'-2' or 2'-1' or 1'-1, while face type V can be reflected in the planes of its edge pairs 2-2 or 2-1 or 1-1'. After performing these operations, all possible fundamental domains associated with a given cell are obtained, in other words, all minimal-cardinality periods are generated.

In this work, the fundamental domains assigned to each cell are denoted by the cell label together with the symbols A/B/C/D, according to their enumeration. In Figures 17-21, only the A-type fundamental domains are illustrated, and their symmetry groups are analysed; in this context, differences in the z-direction are also considered. Figure 11 displays the A-type domains next to the corresponding cells. Figure 12 presents the groups of fundamental domains together with the cells assigned to them. Altogether, eight such groups were identified; these are distinguished solely by the edge structure of the curved face, considering all base-cell edges appearing on the curved face of the fundamental domain. Figures 22-25 display all fundamental domains associated with the 26 cells.

■

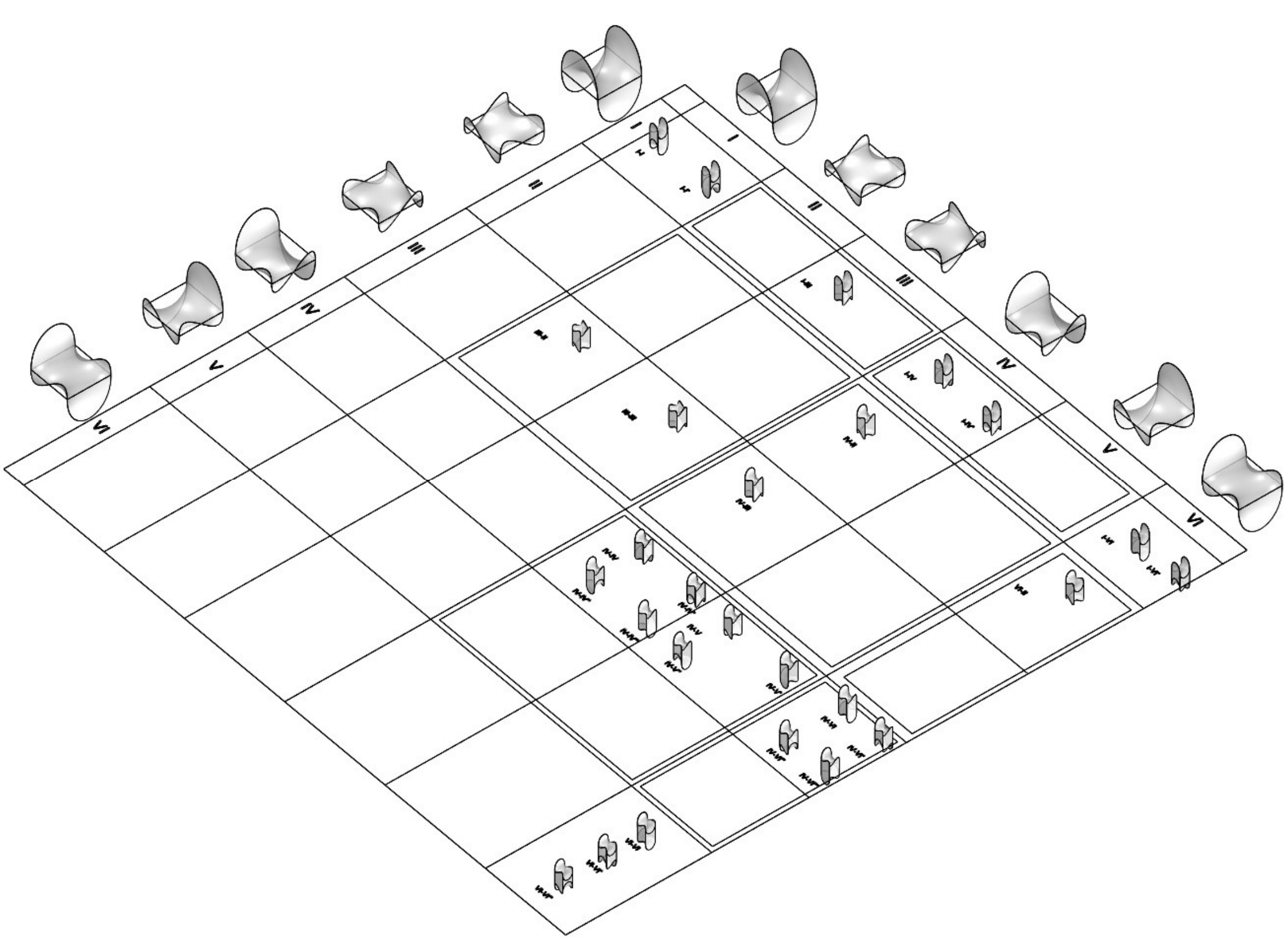

FIGURE 10. Graphical illustration of Table 2. 26 soft cells generated from the cubic lattice, specified by curved face types with the notations used in Lemma 1 and Figure 8.

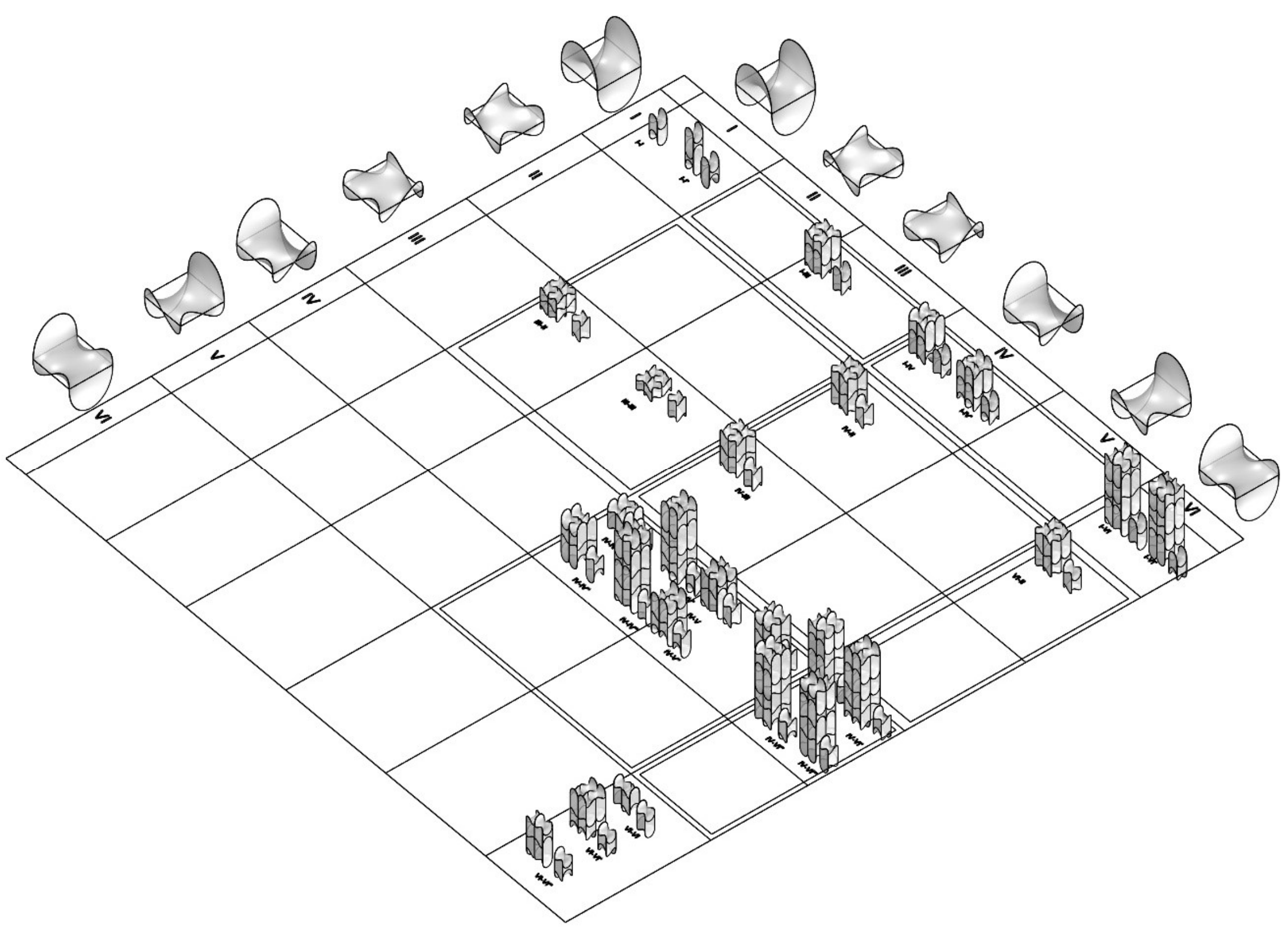

FIGURE 11. Graphical illustration of Table 2, supplemented with the corresponding fundamental domains (minimal sets that can be translated to fill the space). 26 soft cells generated from the cubic lattice, specified by curved face types and their type A fundamental domains using the notations of Theorem 5.

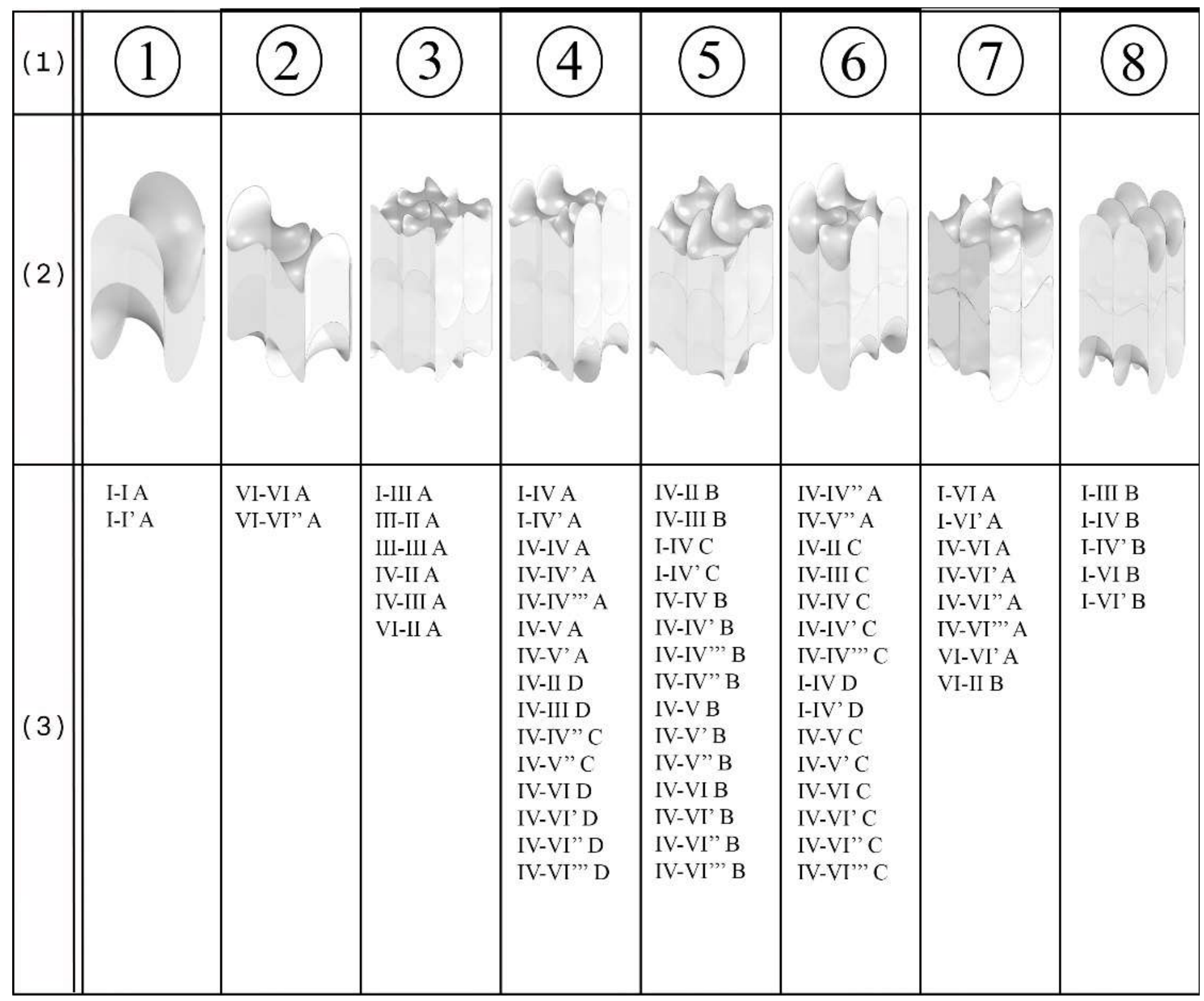

| (1) | ① | ② | ③ | ④ | ⑤ | ⑥ | ⑦ | ⑧ |
|---|---|---|---|---|---|---|---|---|
| (2) | | | | | | | | |
| (3) | I-I A<br>I-I' A | VI-VI A<br>VI-VI" A | I-III A<br>III-II A<br>III-III A<br>IV-II A<br>IV-III A<br>VI-II A | I-IV A<br>I-IV' A<br>IV-IV A<br>IV-IV' A<br>IV-IV''' A<br>IV-V A<br>IV-V' A<br>IV-II D<br>IV-III D<br>IV-IV" C<br>IV-V" C<br>IV-VI D<br>IV-VI' D<br>IV-VI" D<br>IV-VI''' D | IV-II B<br>IV-III B<br>I-IV C<br>I-IV' C<br>IV-IV B<br>IV-IV' B<br>IV-IV''' B<br>IV-IV" B<br>IV-V B<br>IV-V' B<br>IV-V" B<br>IV-VI B<br>IV-VI' B<br>IV-VI" B<br>IV-VI''' B | IV-IV" A<br>IV-V" A<br>IV-II C<br>IV-III C<br>IV-IV C<br>IV-IV' C<br>IV-IV''' C<br>I-IV D<br>I-IV' D<br>IV-V C<br>IV-V' C<br>IV-VI C<br>IV-VI' C<br>IV-VI" C<br>IV-VI''' C | I-VI A<br>I-VI' A<br>IV-VI A<br>IV-VI' A<br>IV-VI" A<br>IV-VI''' A<br>VI-VI' A<br>VI-II B | I-III B<br>I-IV B<br>I-IV' B<br>I-VI B<br>I-VI' B |

FIGURE 12. All fundamental domains (minimal sets that can be translated to fill the space) classified into 8 groups according to their lattice symmetry (according to the geometry of the two opposite curved surfaces). Row (1): notation of lattice symmetry group (1,2,3,4,5,6,7,8). Row (2): a representative fundamental domain with the given lattice symmetry. Row (3): list of fundamental domains with the given lattice symmetry using the notations of Theorem 5.

**s**

| Cell | Fundamental domain | Symmetry group | Order | AC/C |
|---|---|---|---|---|
| I–I | 1×1×1 | $D_{2d}$ | 8 | AC |
| I–I' | 1×1×2 | $D_{2h}$ | 8 | AC |
| I–III | 2x2x2 | $C_2$ | 2 | C |
| I–IV | 2x2x2 | $C_1$ | 1 | C |
| I–IV' | 2x2x2 | $C_1$ | 1 | C |
| I–VI | 2x2x4 | $C_s$ | 2 | AC |
| I–VI' | 2x2x4 | $C_s$ | 2 | AC |

FIGURE 13. Soft cubes and their parameters (I-I - I-VI'). Column (Cell): name and figure of the cell specified by the curved face types using the notations of Lemma 1. Column (Fundamental domain): the size of the fundamental domain (minimal set that can be translated to fill the space) associated with the given cell in the $x-, y-, z-$directions. Column (Symmetry group): name of the symmetry group associated with the given cell and an illustration ( https://newton.ex.ac.uk/research/qsystems/people/goss/symmetry/Solids.html). Column (Order): the order of the symmetry group (the number of symmetry operations mapping the object onto itself). Column (AC/C): $AC$ or $C$, depending on whether the given cell yields an achiral ($AC$) or a chiral ($C$) tiling.

| Cell | Fundamental domain | Symmetry group | Order | AC/C |
|---|---|---|---|---|
| IV-II | 2x2x2 | $C_1$ | 1 | C |
| VI-II | 2x2x2 | $C_1$ | 1 | C |
| III-II | 2x2x2 | $C_{4h}$ | 8 | AC |
| III-III | 2x2x1 | $C_4$ | 4 | C |
| IV-III | 2x2x2 | $C_1$ | 1 | C |
| IV-IV | 2x2x1 | $C_2$ | 2 | C |
| IV-IV' | 2x2x4 | $C_1$ | 1 | C |

FIGURE 14. Soft cubes and their parameters (IV-II - IV-IV'). Column (Cell): name and figure of the cell specified by the curved face types using the notations of Lemma 1. Column (Fundamental domain): the size of the fundamental domain (minimal set that can be translated to fill the space) associated with the given cell in the $x-, y-, z-$directions. Column (Symmetry group): name of the symmetry group associated with the given cell and an illustration ( https://newton.ex.ac.uk/research/qsystems/people/goss/symmetry/Solids.html). Column (Order): the order of the symmetry group (the number of symmetry operations mapping the object onto itself). Column (AC/C): $AC$ or $C$, depending on whether the given cell yields an achiral ($AC$) or a chiral ($C$) tiling.

| Cell | Fundamental domain | Symmetry group | Order | AC/C |
|---|---|---|---|---|
| IV-IV" | 2x2x2 | $C_1$ | 1 | C |
| IV-IV"' | 2x2x4 | $C_1$ | 1 | C |
| IV-V | 2x2x2 | $C_s$ | 2 | AC |
| IV-V' | 2x2x2 | $C_1$ | 1 | C |
| IV-V" | 2x2x2 | $C_i$ | 2 | AC |
| IV-VI | 2x2x4 | $C_1$ | 1 | C |
| IV-VI' | 2x2x4 | $C_1$ | 1 | C |

FIGURE 15. Soft cubes and their parameters (IV-IV" - IV-VI'). Column (Cell): name and figure of the cell specified by the curved face types using the notations of Lemma 1. Column (Fundamental domain): the size of the fundamental domain (minimal set that can be translated to fill the space) associated with the given cell in the $x-, y-, z-$directions. Column (Symmetry group): name of the symmetry group associated with the given cell and an illustration ( https://newton.ex.ac.uk/research/qsystems/people/goss/symmetry/Solids.html). Column (Order): the order of the symmetry group (the number of symmetry operations mapping the object onto itself). Column (AC/C): $AC$ or $C$, depending on whether the given cell yields an achiral ($AC$) or a chiral ($C$) tiling.

| Cell | Fundamental domain | Symmetry group | Order | AC/C |
|---|---|---|---|---|
| IV-VI" | 2x2x4 | $C_1$ | 1 | C |
| IV-VI"' | 2x2x4 | $C_1$ | 1 | C |
| VI-VI | 2x1x1 | $C_s$ | 2 | AC |
| VI-VI' | 2x2x2 | $C_1$ | 1 | C |
| VI-VI" | 2x1x2 | $C_{2v}$ | 4 | AC |

FIGURE 16. Soft cubes and their parameters (IV-VI" - VI-VI"). Column (Cell): name and figure of the cell specified by the curved face types using the notations of Lemma 1. Column (Fundamental domain): the size of the fundamental domain (minimal set that can be translated to fill the space) associated with the given cell in the $x-, y-, z-$directions. Column (Symmetry group): name of the symmetry group associated with the given cell and an illustration ( https://newton.ex.ac.uk/research/qsystems/people/goss/symmetry/Solids.html). Column (Order): the order of the symmetry group (the number of symmetry operations mapping the object onto itself). Column (AC/C): $AC$ or $C$, depending on whether the given cell yields an achiral ($AC$) or a chiral ($C$) tiling.

| Cell | Fundamental domain (A) | Symmetry group | Order |
|---|---|---|---|
| I–I | 1×1×1 | $D_{2d}$ | 8 |
| I–I' | 1×1×2 | $C_{2v}$ | 4 |
| I–III | 2x2x2 | $D_{2d}$ | 8 |
| I–IV | 2x2x2 | $D_{2d}$ | 8 |
| I–IV' | 2x2x2 | $D_{2d}$ | 8 |

FIGURE 17. Fundamental domains and their parameters (I-I - I-IV'). Column (Cell): name of the cell specified by the curved face types using the notations of Lemma 1. Column (Fundamental domain (A)): illustration and the size of the fundamental domain (minimal set that can be translated to fill the space) associated with the given cell in the $x-, y-, z-$directions. Column (Symmetry group): name of the symmetry group associated with the given cell and an illustration ( https://newton.ex.ac.uk/research/qsystems/people/goss/symmetry/Solids.html). Column (Order): the order of the symmetry group (the number of symmetry operations mapping the object onto itself).

| Cell | Fundamental domain (A) | Symmetry group | Order |
|---|---|---|---|
| I–VI | 2x2x4 | $C_{2v}$ | 4 |
| I–VI’ | 2x2x4 | $C_{2v}$ | 4 |
| IV-II | 2x2x2 | $D_{2d}$ | 8 |
| VI-II | 2x2x2 | $D_{2d}$ | 8 |
| III-II | 2x2x2 | $D_{2d}$ | 8 |
| III-III | 2x2x1 | $C_{2v}$ | 4 |

FIGURE 18. Fundamental domains and their parameters (I-VI - III-III). Column (Cell): name of the cell specified by the curved face types using the notations of Lemma 1. Column (Fundamental domain (A)): illustration and the size of the fundamental domain (minimal set that can be translated to fill the space) associated with the given cell in the $x-, y-, z-$directions. Column (Symmetry group): name of the symmetry group associated with the given cell and an illustration ( https://newton.ex.ac.uk/research/qsystems/people/goss/symmetry/Solids.html). Column (Order): the order of the symmetry group (the number of symmetry operations mapping the object onto itself).

| Cell | Fundamental domain (A) | Symmetry group | Order |
|---|---|---|---|
| IV-III | 2x2x2 | $D_{2d}$ | 8 |
| IV-IV | 2x2x1 | $C_{2v}$ | 4 |
| IV-IV' | 2x2x4 | $C_{2v}$ | 4 |
| IV-IV" | 2x2x2 | $D_{2d}$ | 8 |
| IV-IV"' | 2x2x4 | $C_{2v}$ | 4 |
| IV-V | 2x2x2 | $D_{2d}$ | 8 |

FIGURE 19. Fundamental domains and their parameters (IV-III - IV-V). Column (Cell): name of the cell specified by the curved face types using the notations of Lemma 1. Column (Fundamental domain (A)): illustration and the size of the fundamental domain (minimal set that can be translated to fill the space) associated with the given cell in the $x-, y-, z-$directions. Column (Symmetry group): name of the symmetry group associated with the given cell and an illustration ( https://newton.ex.ac.uk/research/qsystems/people/goss/symmetry/Solids.html). Column (Order): the order of the symmetry group (the number of symmetry operations mapping the object onto itself).

| Cell | Fundamental domain (A) | Symmetry group | Order |
|---|---|---|---|
| IV-V’ | 2x2x2 | $D_{2d}$ | 8 |
| IV-V” | 2x2x2 | $D_{2d}$ | 8 |
| IV-VI | 2x2x4 | $C_{2v}$ | 4 |
| IV-VI’ | 2x2x4 | $C_{2v}$ | 4 |
| IV-VI” | 2x2x4 | $C_{2v}$ | 4 |
| IV-VI”’ | 2x2x4 | $C_{2v}$ | 4 |

FIGURE 20. Fundamental domains and their parameters (IV-V’ - IV-VI’’’). Column (Cell): name of the cell specified by the curved face types using the notations of Lemma 1. Column (Fundamental domain (A)): illustration and the size of the fundamental domain (minimal set that can be translated to fill the space) associated with the given cell in the $x-, y-, z-$directions. Column (Symmetry group): name of the symmetry group associated with the given cell and an illustration ( https://newton.ex.ac.uk/research/qsystems/people/goss/symmetry/Solids.html). Column (Order): the order of the symmetry group (the number of symmetry operations mapping the object onto itself).

| Cell | Fundamental domain (A) | Symmetry group | Order |
|---|---|---|---|
| VI-VI | 2x1x1 | $C_{2v}$ | 4 |
| VI-VI' | 2x2x2 | $D_{2d}$ | 8 |
| VI-VI'' | 2x1x2 | $C_{2v}$ | 4 |

FIGURE 21. Fundamental domains and their parameters (VI-VI - VI-VI''). Column (Cell): name of the cell specified by the curved face types using the notations of Lemma 1. Column (Fundamental domain (A)): illustration and the size of the fundamental domain (minimal set that can be translated to fill the space) associated with the given cell in the $x-, y-, z-$directions. Column (Symmetry group): name of the symmetry group associated with the given cell and an illustration ( https://newton.ex.ac.uk/research/qsystems/people/goss/symmetry/Solids.html). Column (Order): the order of the symmetry group (the number of symmetry operations mapping the object onto itself).

| Cell name | Fundamental domains | | | |
|---|---|---|---|---|
| | A | B | C | D |
| I-I | ① | | | |
| I-I' | ① | | | |
| I-III | ③ | ⑧ | | |
| I-IV | ④ | ⑧ | ⑤ | ⑥ |
| I-IV' | ④ | ⑧ | ⑤ | ⑥ |

FIGURE 22. Fundamental domains (A, B, C, D) assigned to the soft cubes with their lattice symmetry groups I-I – I-IV', 68 in total. Column (Cell name): name of the cell using the notations of Lemma 1. Columns (A)-(D): illustration and lattice symmetry group of the A/B/C/D fundamental domains respectively (minimal set that can be translated to fill the space) associated with the given cell using the notations of Figure 12 (1,2,3,4,5,6,7,8).

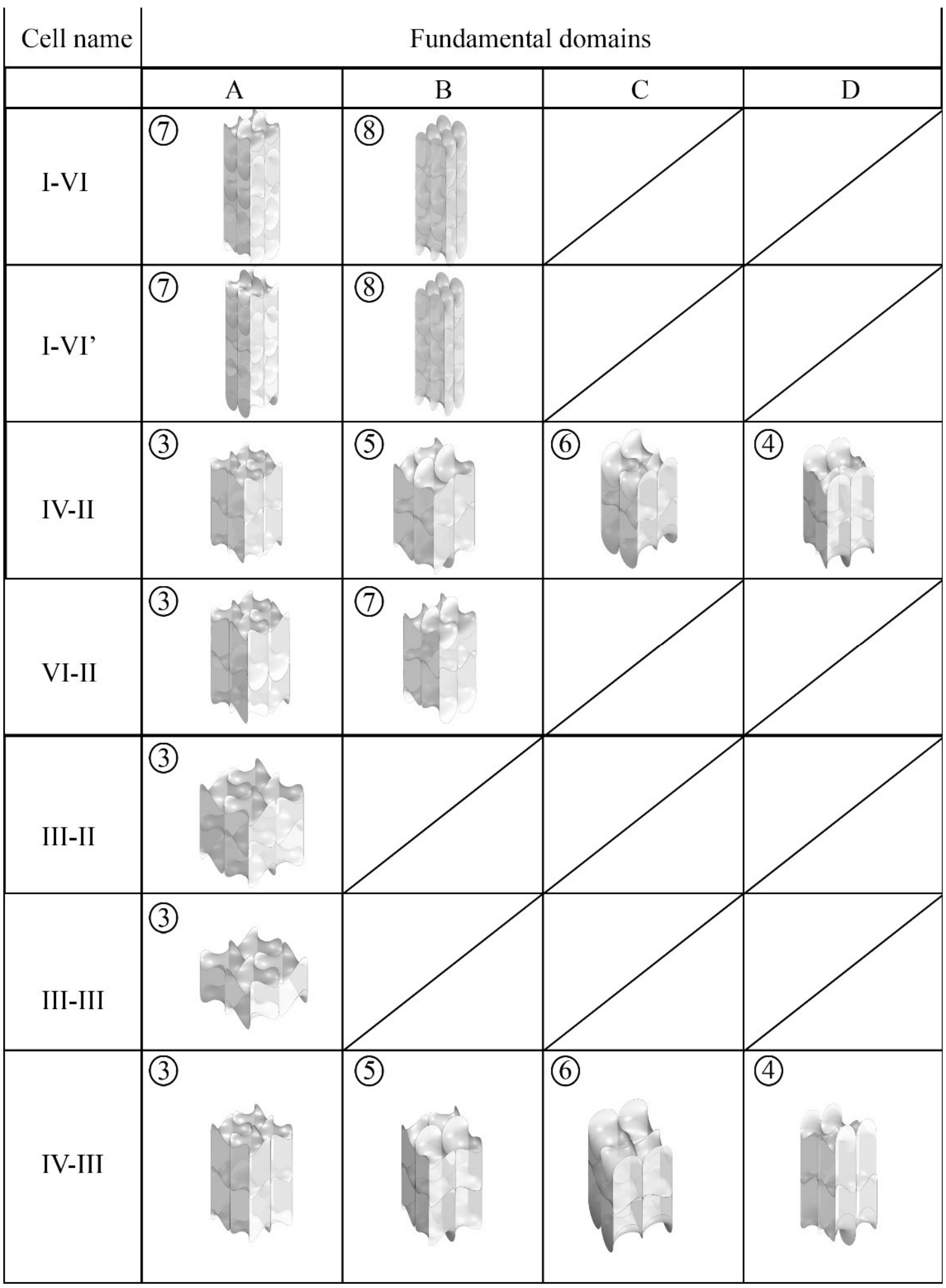

FIGURE 23. Fundamental domains (A, B, C, D) assigned to the soft cubes with their lattice symmetry groups I-VI – IV-III, 68 in total. Column (Cell name): name of the cell using the notations of Lemma 1. Columns (A)-(D): illustration and lattice symmetry group of the A/B/C/D fundamental domains respectively (minimal set that can be translated to fill the space) associated with the given cell using the notations of Figure 12 (1,2,3,4,5,6,7,8).

| Cell name | Fundamental domains | | | |
|---|---|---|---|---|
| | A | B | C | D |
| IV-IV | ④ | ⑤ | ⑥ | |
| IV-IV’ | ④ | ⑤ | ⑥ | |
| IV-IV’’ | ⑥ | ⑤ | ④ | |
| IV-IV’’’ | ④ | ⑤ | ⑥ | |
| IV-V | ④ | ⑤ | ⑥ | |
| IV-V’ | ④ | ⑤ | ⑥ | |
| IV-V’’ | ⑥ | ⑤ | ④ | |

FIGURE 24. Fundamental domains (A, B, C, D) assigned to the soft cubes with their lattice symmetry groups IV-IV – IV-V’’, 68 in total. Column (Cell name): name of the cell using the notations of Lemma 1. Columns (A)-(D): illustration and lattice symmetry group of the A/B/C/D fundamental domains respectively (minimal set that can be translated to fill the space) associated with the given cell using the notations of Figure 12 (1,2,3,4,5,6,7,8).

| Cell name | Fundamental domains | | | |
|---|---|---|---|---|
| | A | B | C | D |
| IV-VI | ⑦ | ⑤ | ⑥ | ④ |
| IV-VI’ | ⑦ | ⑤ | ⑥ | ④ |
| IV-VI” | ⑦ | ⑤ | ⑥ | ④ |
| IV-VI”’ | ⑦ | ⑤ | ⑥ | ④ |
| VI-VI | ② | | | |
| VI-VI’ | ⑦ | | | |
| VI-VI” | ② | | | |

FIGURE 25. Fundamental domains (A, B, C, D) assigned to the soft cubes with their lattice symmetry groups IV-VI – VI-VI”, 68 in total. Column (Cell name): name of the cell using the notations of Lemma 1. Columns (A)-(D): illustration and lattice symmetry group of the A/B/C/D fundamental domains respectively (minimal set that can be translated to fill the space) associated with the given cell using the notations of Figure 12 (1,2,3,4,5,6,7,8).

From the complete classification it follows that there exists no soft space-filling tiling that is combinatorially equivalent to the cube and at the same time preserves its full symmetry group. Among the soft space-filling tilings obtainable from the cubic lattice, the richest symmetry groups are of order eight. These include the group $D_{2d}$ (realized by the I–I cell), $\mathrm{D_{2h}}$ (realized by the I–I' cell) and the group $\mathrm{C_{4h}}$ (realized by the III–II cell).

| Crystal family | Crystal system | Lattice system | Required symmetries of the point group | Point groups | Space groups | Bravais lattices |
|---|---|---|---|---|---|---|
| **Triclinic** | Triclinic | Triclinic | None | 2 | 2 | 1 |
| **Monoclinic** | Monoclinic | Monoclinic | 1 twofold axis of rotation or 1 mirror plane | 3 | 13 | 2 |
| **Orthorhombic** | Orthorhombic | Orthorhombic | 3 twofold axes of rotation or 1 twofold axis of rotation and 2 mirror planes | 3 | 59 | 4 |
| **Tetragonal** | Tetragonal | Tetragonal | 1 fourfold axis of rotation | 7 | 68 | 2 |
| **Hexagonal** | Trigonal | Rhombohedral | 1 threefold axis of rotation | 5 | 7 | 1 |
| | | Hexagonal | 1 threefold axis of rotation | 5 | 18 | 1 |
| | Hexagonal | Hexagonal | 1 sixfold axis of rotation | 7 | 27 | 1 |
| **Cubic** | Cubic | Cubic | 4 threefold axes of rotation | 5 | 36 | 3 |
| 6 | 7 | 7 | **Total** | 32 | 230 | 14 |

FIGURE 26. Classification of crystal families. [4]

| | | | | | | | | |
|---|---|---|---|---|---|---|---|---|
| triclinic | pedial | $C_1$ | 1 | 11 | $[\,]^+$ | enantiomorphic polar | 1 | trivial $\mathbb{Z}_1$ |
| | pinacoidal | $C_i$ ($S_2$) | $\bar{1}$ | 1x | $[2,1^+]$ | centrosymmetric | 2 | cyclic $\mathbb{Z}_2$ |
| monoclinic | sphenoidal | $C_2$ | 2 | 22 | $[2,2]^+$ | enantiomorphic polar | 2 | cyclic $\mathbb{Z}_2$ |
| | domatic | $C_s$ ($C_{1h}$) | m | *11 | [ ] | polar | 2 | cyclic $\mathbb{Z}_2$ |
| | prismatic | $C_{2h}$ | 2/m | 2* | $[2,2^+]$ | centrosymmetric | 4 | Klein four $\mathbb{V} = \mathbb{Z}_2 \times \mathbb{Z}_2$ |
| orthorhombic | rhombic-disphenoidal | $D_2$ (V) | 222 | 222 | $[2,2]^+$ | enantiomorphic | 4 | Klein four $\mathbb{V} = \mathbb{Z}_2 \times \mathbb{Z}_2$ |
| | rhombic-pyramidal | $C_{2v}$ | mm2 | *22 | [2] | polar | 4 | Klein four $\mathbb{V} = \mathbb{Z}_2 \times \mathbb{Z}_2$ |
| | rhombic-dipyramidal | $D_{2h}$ ($V_h$) | mmm (2/m 2/m 2/m) | *222 | [2,2] | centrosymmetric | 8 | $\mathbb{V} \times \mathbb{Z}_2$ |
| tetragonal | tetragonal-pyramidal | $C_4$ | 4 | 44 | $[4]^+$ | enantiomorphic polar | 4 | cyclic $\mathbb{Z}_4$ |
| | tetragonal-disphenoidal | $S_4$ | $\bar{4}$ | 2x | $[2^+,2]$ | non-centrosymmetric | 4 | cyclic $\mathbb{Z}_4$ |
| | tetragonal-dipyramidal | $C_{4h}$ | 4/m | 4* | $[2,4^+]$ | centrosymmetric | 8 | $\mathbb{Z}_4 \times \mathbb{Z}_2$ |
| | tetragonal-trapezohedral | $D_4$ | 422 | 422 | $[2,4]^+$ | enantiomorphic | 8 | dihedral $\mathbb{D}_8 = \mathbb{Z}_4 \rtimes \mathbb{Z}_2$ |
| | ditetragonal-pyramidal | $C_{4v}$ | 4mm | *44 | [4] | polar | 8 | dihedral $\mathbb{D}_8 = \mathbb{Z}_4 \rtimes \mathbb{Z}_2$ |
| | tetragonal-scalenohedral | $D_{2d}$ ($V_d$) | $\bar{4}$2m | 2*2 | $[2^+,4]$ | non-centrosymmetric | 8 | dihedral $\mathbb{D}_8 = \mathbb{Z}_4 \rtimes \mathbb{Z}_2$ |
| | ditetragonal-dipyramidal | $D_{4h}$ | 4/mmm (4/m 2/m 2/m) | *422 | [2,4] | centrosymmetric | 16 | $\mathbb{D}_8 \times \mathbb{Z}_2$ |

FIGURE 27. Triclinic, monoclinic, orthorhombic, and tetragonal crystal systems. [4]

## 5. SUMMARY

Possible softenings of the cubic lattice were examined. Using the algorithm shown in [3], we showed that 12 solutions are available for the complete sets of softening equations, if the half-tangent vectors of the edges are restricted to lattice directions, and the edges of the tiles are planar. We proved that one, and only one solution of the complete sets of softening equations provide the softness of all tiles at each node. (This solution appears in 4 equivalent solutions among the 12.) We showed that parallel soft tilings that fill the space without gaps and overlaps and partially agree to second order with the cubic lattice have prismatic structure. We determined the possible curved edge and face configurations with the edge half-tangent vectors of soft cubes. We proved that there are exactly 26 parallel soft tilings which partially agree to second order with the cubic lattice, considering the one published in [1]. To this end, we presented an algorithm (Algorithm 2) with its implementation in Python, which beyond identifying the 26 distinct geometries, determines which cells tile the space by themselves or by their rotations (achiral tiling), and which require a mirror copy to tile the space (chiral tiling). We assigned each cell its corresponding symmetry group. We showed that, in total, the cells tile space in 68 distinct ways by repetition of minimal sets (fundamental domains). The fundamental domains were classified into 8 groups according to their lattice symmetry.

**ACKNOWLEDGEMENTS**

The author wishes to thank Prof. Gábor Domokos and Prof. Ákos G. Horváth for their valuable insights and support during the preparation of this manuscript.

KINGA KOCSIS, DEPARTMENT OF MORPHOLOGY AND GEOMETRIC MODELLING and HUN-REN-BME MORPHODYNAMICS RESEARCH GROUP, BUDAPEST UNIVERSITY OF TECHNOLOGY AND ECONOMICS, Műegyetem Rkp. 3., K220, Budapest 1111, Hungary

Email address: kocsis.kina@gmail.com

**APPENDIX**

The Python code of Algorithm 2 is the following:

```
from math import*
#define edges
A=[0,0]
B=[1,0]
C=[1,1]
D=[0,1]
ELEK=[A,B,C,D]

#define every possible curved face in every orientation
LAPOK=[]
for i in ELEK:
   for j in ELEK:
      for k in ELEK:
         for l in ELEK:
            if (i[1]!= j[0] and j[1] != k[0] and k[1]!= l[0] and l[1] != i[0]):
               LAPOK.append([i,j,k,l])
#face pairings
CELLAK=[]
for a in range(len(LAPOK)):
   for b in range(len(LAPOK)):
      CELLAK.append((LAPOK[a],LAPOK[b]))
#transformations
#k is the rotation: 0,1,2,3 --> 0, pi/2,pi,3pi/2
def z_rot_lap(lap, k):
   k = k % 4
   return tuple(lap[k:] + lap[:k])
def z_rot_par(par,k):
   lap1, lap2=par
   return(z_rot_lap(lap1,k), z_rot_lap(lap2,k))
#rotations about the y-axis by pi
def y_rot_lap(lap):
   #switch edge orientation
   def trans(el):
      x, y = el
      return (1 - y, 1 - x)
```

```
#switch traversal direction
    return tuple(
        trans(tuple(x))
        for x in [lap[0], lap[3], lap[2], lap[1]]
    )

#face switch
def y_rot_par(par):
    lap1, lap2=par
    return(y_rot_lap(lap2), y_rot_lap(lap1))
#every possible transformation combination
def minden_transzf(par):
    eredmeny = []
    #z rotation 4
    for k in range(4):
        z = z_rot_par(par, k)
        eredmeny.append(z)

        #y rotation
        yz = y_rot_par(z)
        eredmeny.append(yz)

        #z rotation again
        for m in range(4):
            z_yz = z_rot_par(yz, m)
            eredmeny.append(z_yz)

            # y rotation again: Y Z^m Y Z^k
            y_z_yz = y_rot_par(z_yz)
            eredmeny.append(y_z_yz)

    # remove duplicates
    return list(set(eredmeny))

#turn list to tuple
def to_tuple_par(par):
    lap1, lap2 = par
```

```
    return (tuple(tuple(x) for x in lap1),
            tuple(tuple(x) for x in lap2))

#similarity of two pairs
def azonos(par1, par2):
    for t in minden_transzf(par1):
        if t == par2:
            return True
    return False

#distinct geometries
KULONBOZO = []
for i in range(len(CELLAK)):
    par1 = to_tuple_par(CELLAK[i])

    uj = True

    for j in range(len(KULONBOZO)):
        if azonos(par1, KULONBOZO[j]):
            uj = False
            break

    if uj:
        KULONBOZO.append(par1)

print(len(KULONBOZO))

#dictionary for printing
nev = {
    (0,0): "A",
    (1,0): "B",
    (1,1): "C",
    (0,1): "D"
}

def lap_to_str(lap):
    return "".join(nev[tuple(elem)] for elem in lap)
```

```
def par_to_str(par):
    lap1, lap2 = par
    return lap_to_str(lap1) + " - " + lap_to_str(lap2)

#printing
for par in KULONBOZO:
    print(par_to_str(par))

for par in KULONBOZO:
    print(par)
```